\documentclass{article}

\usepackage{graphicx}
\usepackage{latexsym}

\begin{document}

\centerline{\LARGE{A local theory of Quantum Mechanics}}

\centerline{Carlos Lopez}

\centerline{Dept. of Physics and Mathematics, Facultad de Ciencias}
\centerline{UAH, Alcal\'a de Henares , E-28871 (Madrid, SPAIN)}
\centerline{carlos.lopez\@ uah.es     }



\begin{abstract}
It is shown that Quantum Mechanics is ambiguous when predicting
relative frequencies for an entangled system if the measurements of both subsystems
are performed in spatially separated events. This ambiguity gives way to
unphysical consequences: the projection rule could be applied in one or the other
temporal(?) order of measurements (being non local in any case), but symmetry
of the roles of both subsystems would be broken.

An alternative theory is presented in which this ambiguity does not exist.
Observable relative frequencies differ from those of orthodox Quantum Mechanics,
and a {\it gendaken} experiment is proposed to falsify one or the other theory.
  In the alternative theory, each subsystem has an individual state in
its own Hilbert space, and the total system state is direct product (rank one) of both, so
there is no entanglement. Correlation between subsystems appears through a
hidden label that prescribes the output of arbitrary hypothetical measurements.

Measurement is treated as a usual reversible interaction, and this postulate allows to 
determine relative frequencies 
when the value of a magnitude is known without in any way perturbing the system, by 
measurement of 
the correlated companion. 

It is predicted the existence of an
 accompanying  system, the de Broglie wave,
introduced in order to preserve the action reaction principle in
indirect measurements, when there is no interaction of detector and
particle. Some action on the detector, different from the one cause by
a particle, should be observable.

\end{abstract}

\section{Introduction}
\label{intro}

In the author's alternative interpretation of Quantum Mechanics \cite{mio},
  both spinless point particles in phase space and $1/2$ spin
particles in the space of spin states were represented in an enlarged Hilbert
space, where physical states can have simultaneously definite values of two or
more non commuting magnitudes, position and momentum in phase space, two or more
spin directions in spin space. Observable predictions of Quantum Mechanics (the
marginal amplitudes of probability) were reproduced in individual systems, but
the existence of an additional physical ingredient, the de Broglie wave
\cite{dBro}, was proposed in order to preserve the action reaction principle, which seems to be 
violated in Quantum Mechanics for indirect measurements.
The  state of the  particle with respect to a given magnitude was
described through a label $\lambda$, attached to one of the non null components
of the vector state as linear combination of eigenstates of
the  
magnitude (position, momentum, spin in a given direction). For particles
correlated in spin variables,  entanglement in the product of Hilbert  spaces 
of both particles was
replaced by a $\lambda$ correlation prescribing a total null spin,
 and the composite state was direct product of individual ones. 

In the former interpretation there is no scientific (observable) contradiction
with orthodox Quantum Mechanics,
relative frequencies for measurements of individual systems match the known
values. 
The local $\lambda$ description of correlation for spin variables is also an interpretation, 
 Aspect's experiment \cite{Aspect}
with polarisation of photons and Bell's {\it gendaken} experiment \cite {Bell} with spin  are satisfactorily reproduced.

Measurement over an eigenstate (change of state of the apparatus but not
of the measured system) suggest the existence of hidden variables in order to
preserve the action reaction principle. As a new observable system (but not
contradictory with the orthodox theory), it was predicted the existence of a real, physical
wave, interacting with the detector when virtual
paths of a process are discarded because the particle is not detected, and the
projection rule is applied to take into account this fact. Detector and particle
do not interact, but there is an observable change of state in the particle. The suppressed (or phase shift of) wave component would explain this change of state 
in case some measurable effect could be observed on the detector.

The theory presented in this article applies the former ideas, states with
simultaneously definite values of non commuting magnitudes, hidden label
prescribing the output of measurements, $\lambda$ correlation, to a generic
quantum system, state in a Hilbert space, set of self adjoint operators.
While there is no observable distinction with orthodox Quantum Mechanics
for individual systems (up to the undetected de Broglie wave), 
a different table of relative frequencies is obtained for generic correlated
systems and variables, although this difference disappears in particular cases, e.g.,
for spin  and polarisation  variables.
In fact, there is ambiguity in the standard theory for generic cases in the  prediction of the 
distribution of probabilities for measurement events that are spatially separated; 
only if probabilities for both possible temporal orders of measurement coincide
the ambiguity disappears. 

In the proposed Quantum Mechanics with label, the distribution becomes 
different from the orthodox one even when there is well established time order in the measurement events and 
the projection rule should be applied, because in the alternative theory there is no projection of state 
in the system not being measured, and the type
of space time separation between measurements is irrelevant.
For two correlated systems $S_a$ and $S_b$, measurement of system $S_b$ is just a source of information about  outputs of an hypothetical measurement on  system $S_a$.  Measurement is considered a usual Hamiltonian, 
reversible interaction; 
the distribution in the  $S_a$ state of probabilities for the (correlated) variable (the one measured at $S_b$)
is preserved under the real $S_a$ measurement, that is, knowledge of it after measurement 
(real interaction, projection of state) determines the distribution previous to the interaction. This postulate of reversibility, a kind of ``virtual'' (non interacting) measurement,  allows
to obtain predictions of relative frequencies; they match the standard ones 
in absence of projection rule, and determine the statistical correlation with the other system's measurement outputs.

Next section presents the theory in  the general case 
for individual systems.
The previous examples \cite{mio} of phase space for a point like spinless particle and spin
state space are revisited. They are certainly the most relevant; if entanglement 
 is replaced by direct product
of individual states, as described in section $3$, 
every quantum system can be decomposed into its elementary
subsystems (elementary particles). Section
$3$ introduces the description of correlated systems through the $\lambda$
mechanism, and when Born's rule is applied, both in real and ``virtual'' measurements,
we get relative frequencies
to be compared with the orthodox theory. A {\it gendaken} two sit experiment,
with an additional  system 
correlated to the right--left slit variable of the particle, shows that, while in
the standard theory the diffraction pattern disappears for early measurement of
the other system, when it is known before reaching the final screen which slit the particle is coming from,
in this formulation diffraction pattern is preserved, even when the
correlated system's very early measurement event allows to know in advance which slit the particle  will
go  through at the first screen.

\section{Description of a generic quantum system}
\label{descr}

Let $\cal S$ be a quantum system, ${\cal F} = \{A,B,C, \ldots \}$ physical
magnitudes (and their self adjoint representation in the Hilbert space $\cal H$
of the system), $|S>$ a particular quantum state of $\cal S$, unit vector (or
ray) in $\cal H$. We will denote by $a_i$, $b_j$, \ldots,  eigenvalues of $A$,
$B$, etc., and $|a_i>$, 
$|b_j>$ eigenvectors of the shown eigenvalue.  If $|S> = \sum _i z^A_i |a_i>$,
all $a_i$ with non vanishing $z^A_i$ are possible outputs of a measurement of magnitude $A$
on $|S>$. For a given family ${\cal F}$ of not necessarily commuting magnitudes,
the set of  $n$--tuples of {\bf consistently} (in $|S>$) joint
 values for these magnitudes is defined by

\[
{\cal M}_{|S>} = \{ (a_i, b_j, c_k, \ldots) | <a_i|b_j> \neq 0, <a_i|c_k> \neq 0,  \ldots \}
\]
i.e., $a_i$ and $b_j$ do not belong to the same element of ${\cal M}_{|S>}$ if
pure state $|a_i>$ has not 
$b_j$ as possible output of a $B$--measurement (and vice versa). The same
requirement applies to all pairs of values for different magnitudes. 
These values
$(a_i, b_j, c_k, \ldots)$ are formally defined, and can not be jointly and
consistently measured on $|S>$ for non commuting magnitudes. We will denote by
$\pi _A : {\cal M}_{|S>}  \to {\cal E}_A = \{ a_i \}$ the projection over the
set of $A$ eigenvalues, $\pi _B$ over ${\cal E}_B$, etc.

In orthodox Quantum Mechanics, it is on ${\cal E}_A$, 
${\cal E}_B$, \ldots, and not on ${\cal M}_{|S>}$, where a distribution of
amplitudes of probability is defined, and this is just the quantum state
$|S>$, with corresponding amplitudes $z^A_i$, $z^B_j$, etc.
If $M_A(|S>)$ represents the output of an $A$ measurement performed on $|S>$,
$M_A(|S>) \in {\cal E}_A$, the distribution of probabilities for these outputs is
$P(a_i) = |<a_i|S>|^2$. After measurement, the new state of the system is
$|a_i>$ when $M_A(|S>) = a_i$. 

In the usual interpretation, a state $|S> \neq |a_i>$ has no definite value of
magnitude $A$ (but it can have forbidden values $z^A_i = 0$), and it is at
measurement when a precise value is achieved. We could put a label on the
element of ${\cal E}_A$ output of measurement. In alternative interpretations,
the label exists (but hidden) before measurement is performed, and determines
its output. This hypothesis is introduced in order to interpret measurement as a usual
interaction, and the projection rule as both a prescription of the new state
(after interaction with the apparatus) and a probabilistic rule over a family of non
identical physical systems. 

Obviously, a hidden label without any other modification in
the mathematical machinery, is just an interpretative matter. But, if the label
is present (an element of reality \cite{EPR}) before measurement, the pure state
$|S>$ must be a sample of non identical physical systems, with the label in 
different positions. Moreover, while we  put the
label generated at measurement once at a time for each successive measurement of non commuting
observables, if it is always present there must be all the time a label at each ${\cal
E}_A$, ${\cal E}_B$, etc. ready to determine the output of an arbitrary
measurement. Taking into account the previous consistency requirement,
the label will be  an element  $\lambda \in {\cal M}_{|S>}$. 

Let us consider  the pure quantum state  as a sample, family of 
physical systems, $|S> = \{ |S; \lambda > ; \lambda \in {\cal M}_{|S>} \}$, all
of them represented  by the same state vector $|S>$ in $\cal H$,
but each one with different label. The label completely (for all considered magnitudes)
determines the output of arbitrary measurements of the system. While $M_A(|S>)
\in {\cal E}_A$ is probabilistic, $M_A(|S; \lambda>) = \pi _A (\lambda )$ is
fixed for an hypothetically known label. Obviously, the probabilistic character of the theory does not
disappear, it is just displaced from each ${\cal E}_A$, ${\cal E}_B$, \ldots,
onto ${\cal M}_{|S>}$; in this notation, $P(\pi _A (\lambda )) = |<\pi _A(\lambda )|S>|^2$
replaces 
$P(a_i) =$ $|<a_i|S>|^2$ when applied to the statistical ``universe'' $|S> = \{ |S; \lambda >\}$. The
former relation means:
``there is probability $P(\pi _A (\lambda ) = a_i)$ that a randomly chosen
element $|S; \lambda >$ of $|S>$ has $\pi _A (\lambda ) = a_i$''.  

We can ask ourselves if there is a distribution of probabilities in ${\cal
M}_{|S>}$, $P(\lambda )$,  such that

\begin{enumerate}

\item Marginal probabilities match the quantum theory values

\[
\sum _{\lambda \ni \pi _A(\lambda ) = a_i} P(\lambda ) = P(\pi _A (\lambda )) =
P(a_i) = |z^A_i|^2
\]
The previous sum is restricted to labels fulfilling the given condition.
A similar condition must hold for all considered magnitudes $B$, $C$, \ldots

\item The condition of consistency, contained in the definition of ${\cal M}_{|S>}$, which is a subset of 
${\cal E}_A \times {\cal E}_B \times \cdots$

\[
P(\lambda ) = 0 \hskip 1 truecm {\rm if} \hskip 1 truecm <\pi _A (\lambda )|\pi
_B (\lambda > = 0 
\,\, {\rm or } \,\ldots
\]
i.e., $P(\lambda )$ vanishes if  some pair of projected values are incompatible.

\item Positivity, $P(\lambda ) \geq 0$. From $1$ and $3$ we get $P(\lambda ) = 0$ when
$P(\pi _A (\lambda )) = 0$ for some $A$.

\end{enumerate}

It is well known that the third requirement can not be accomplished in generic cases, as
Bell's type inequalities theorems prove \cite {Bell}. The first and second
requirements are linear equations, generically with more unknowns than
equations, and have usually many solutions. Symmetry considerations can
establish a preferred solution. For two magnitudes it is easy to find a general
solution, with notation $W$ instead of $P$ because of its quasi probability
(weight, or Wigner \cite{Wigner}) character,

\[
W(a_i, b_j) = \frac {1}{2} \left( z^{*A}_i z^B_j <a_i|b_j> +  z^{*B}_j z^A_i
<b_j|a_i>\right)
\]
from where 

\[
\sum _j W(a_i, b_j) =  \frac {1}{2}\left( z^{*A}_i <a_i|\sum _j z^B_j b_j> +
z^A_i (\sum _j z^{*B}_j <b_j)|a_i>\right) = 
\]

\[
= \frac {1}{2}\left(
z^{*A}_i <a_i|S> + z^A_i <S|a_i>\right) = z^{*A}_i z^A_i = P(a_i)
\]

For example,  Wigner's quasi probability distribution in phase space is obtained
from

\[
1= 
\frac {1}{2} \left(
\int dx_1 \Psi ^*(x_1) <x_1| \int dp \xi (p) |p> + \int dp \xi ^*(p) <p| \int
dx_1 \Psi (x_1) |x_1>\right)
\]

\[
 = \frac {1}{2} \left(
\int dx_1  \int dp \int dx_2 \Psi ^*(x_1)  \Psi (x_2) e^{i/\hbar p(x_1 - x_2)} 
+  {\rm conjugate} \right) 
\]
and, after change of variables $x_1 = x + s/2$, $x_2 = x - s/2$, we find the well
known Wigner's kernel $W(x, p)$,

\[
1 = \int dx  \int dp [\int ds \Psi ^*(x + s/2)  \Psi (x - s/2) e^{i/\hbar ps}] =
\int dx  \int dp W(x,p)
\]

There is no physical interpretation for these weights; moreover, they are
se\-con\-da\-ry ingredients, built from the amplitudes. Although there is
neither a clear physical interpretation of amplitudes of probability, the path
integral formalism \cite{Path} suggest its association with some physical wave,
the de Broglie wave \cite{mio}
of which phases are added in a diffusion, propagation process with the
typical interference phenomenon. We can ask ourselves if there is a
distribution $Z(\lambda )$, of amplitudes of probability in ${\cal M}_{|S>}$, 
representation of the wave, and
such that

\begin{enumerate}

\item  Marginal amplitudes match the quantum theory

\[
\sum _{\lambda \ni \pi _A(\lambda ) = a_i} Z(\lambda ) = z^A_i
\]
for each magnitude $A$, $B$, \ldots

\item Consistency in ${\cal M}_{|S>}$

\[
Z(\lambda ) = 0 \hskip 1 truecm {\rm if} \hskip 1 truecm <\pi _A (\lambda )|\pi
_B (\lambda > = 0 
\,\, {\rm or } \,\ldots
\]
i.e, whenever some $<a_i|b_j> = 0$.

\item No positivity requirement, amplitudes are not even real numbers!

\end{enumerate}

The third point allows (does not prevent) the existence of consistent solutions,
contrarily to the probabilities distribution problem. Again, they are linear equations with
many solutions in general. However, only one solution
would have physical meaning if these amplitudes (i.e., relative frequencies
through Born's rule) were observable. 
For two magnitudes $A$ and $B$, with amplitude $Z_{i j} = Z(a_i, b_j)$ and associated joint
probability (always positive!) 
$|Z_{i j}|^2$, the relative frequencies can not be observed if $A$ and $B$ do not
commute. Measurement of $A$ necessarily modifies  $\pi _B(\lambda )$,
 and vice versa. We  can only observe $P(a_i)=|z^A_i|^2$, and then
$P(b_j|a_i)=|<b_j|a_i>|^2$ on the new state $|a_i>$ after projection. Similarly,
$P(b_j)= |z^B_j|^2$ for $B$ measurement, and $P(a_i|b_j)$ (which equals
$P(b_j|a_i)$) on the new state $|b_j>$ after projection.  $P(a_i)P(b_j|a_i) \neq
P(b_j)P(a_i|b_j)$ unless $|z^A_i|^2 = |z^B_j|^2$.

Let us consider a simple measurement of magnitude $A$,
performed on a known eigenstate $|a_i>$ of $A$. Before measurement, the distribution of
amplitudes of probability for $B$ 
is $<b_j|a_i>$. The label in a particular system $|a_i, \lambda >$
 of the given state is an unknown $b_j =\pi _B(\lambda )$ together with the known $a_i$, 
$\lambda = (a_i, b_j)$. In this case we know the 
distribution of amplitudes $Z_{i j} = Z(\lambda
=(a_i, b_j)) = <b_j|a_i>$ (or $Z_{i' j} = \delta _{i' i} <b_j | a_i>$) and  probabilities
$|<b_j|a_i>|^2$, according to Born's rule, before measurement. This is the
distribution to be found if $B$ were measured in a sample of systems (identical in the standard interpretation,  with different $\lambda$ here). As we measure $A$, the
output is obviously $a_i$, and there is no change of pure quantum state. The new distribution 
matches the initial one.

If measurement is
a usual Hamiltonian interaction with Hamiltonian $H(A, Y)$, depending on the 
measured magnitude $A$ and some variable $Y$ of the apparatus, 
action over the measurement system (new state of the pointer)
is accompanied by
a reaction over $|a_i, \lambda >$, i.e., an output system $|a_i,\lambda '>$
of the same quantum state, both
$\pi _B(\lambda ) = b_j$ and $\pi _B(\lambda ') = b'_j$ unknown,  with $b'_j \neq b_j$. 
\footnote{In indirect measurements there is no physical interaction of particle, the label,
with the detector. Therefore, the label does not change. It is the de Broglie wave that physically interacts,
and either it is suppressed by the obstacle or there is a phase shift. This 
 phenomenon, necessary to maintain the action reaction principle, has not been observed.}
The relevant point
is that the distribution  $Z(\lambda ')$ equals  $Z(\lambda )$, according to quantum rules.

Notice that 
$\pi _A (\lambda ) = a_i = \pi _A(\lambda ')$ is preserved  in an {\bf arbitrary} 
measurement of magnitude $A$, from an input $|S, \lambda>$  with $\pi _A(\lambda ) = a_i$ unknown before measurement ($|S> \neq |a_i>$ in a generic case), onto the output $|a_i, \lambda '>$: the ideal Hamiltonian of interaction
commutes with $A$. By reversibility of  Hamiltonian interactions,
when $A$ is measured the statistical distribution of the $b_j$ is not modified,
although the label changes from  $\pi _B(\lambda ) = b_j$ to $\pi _B(\lambda ') = b'_j \neq b_j$ at each particular measurement
event. That is, as we get an output $|a_i, \lambda '>$ with known distribution $<b_j|a_i>$
for the $b_j$, the same distribution is  the 
{\bf initial} one in state $|S, \lambda>$, a state of which we 
know (a posteriori) that it fulfilled $\pi _A(\lambda )= a_i = \pi _A(\lambda ')$.

We {\bf postulate} that the conditional distribution $Z_{i j} = Z(a_i, b_j)$ for a state $|S, \lambda>$ 
 with $\pi _A(\lambda ) = a_i$ matches the distribution $<b_j|a_i>$
of the eigenstate $|a_i>$. For the joint distribution of amplitudes of probability

\[
\frac {|Z_{i j}|^2}{\sum _{j'} |Z_{i j'}|^2} = |<b_j|a_i>|^2
\]
Similarly, $\pi _B(\lambda ) = b_j$ conditional distribution for $Z_{i j}$ matches
$<b_j|a_i>$,

\[
\frac {|Z_{i j}|^2}{\sum _{i'} |Z_{i' j}|^2} = |<b_j|a_i>|^2
\]
from where we find

\[
\sum ^N_{j'} |Z_{i j'}|^2 = \sum ^N_{i'} |Z_{i' j}|^2 = \frac {1}{N}
\]
in a normalised total state $\sum _{ i' j'} |Z_{i j'}|^2 = 1$.

An initial measurement of $A$ has marginal distribution of amplitudes and probabilities

\[
z^A_i = \sum _{j'} Z_{i j'} \quad P(a_i) = |z^A_i|^2 = |\sum _{j'} Z_{i j'}|^2
\]
or, equivalently,

\begin{equation}
\label{pai}
P(a_i) = |\sum _j z^B_j <a_i | b_j>|^2
\end{equation}
On the other hand, an initial measurement of $B$ gives

\[
P(b_j) = |z^B_j|^2 = |\sum _{i'} Z_{i' j}|^2
\]
and, if  later on we measure $A$ on output states $|b_j>$, we find

\[
P(a_i| b_j) = |<a_i|b_j>|^2 \quad P(b_j)P(a_i | b_j) = |z^B_j|^2|<a_i|b_j>|^2
\]
with marginal probability 

\begin{equation}
\label{ppai}
P'(a_i) = \sum _j P(b_j)P(a_i | b_j) = \sum _j |z^B_j <a_i | b_j>|^2
\end{equation}

This is a well know result: there is interference of all $b_j$ components in $z^A_i$ and $P(a_i)$, while there is a
sum of classical independent probabilities in $P'(a_i)$. But, contrarily to the usual interpretation, interference at $P(a_i)$ does not contradict an assignment of (hidden) $b_j$ values to the initial state. The label contains a 
particular $b_j$ value, but the associated de Broglie wave, represented by the orthodox quantum state,  has all $b_j$ components.

The previous postulate  allows to find an essentially unique solution for $Z$s; the whole information is contained
in the orthodox quantum state. Its
physical foundation is time reversibility of dynamical evolution, including measurement. At all times, not
just under measurement, the label is attached to some
$n$--tuple of values (with some kind of stochastic time evolution) of the physical magnitudes. We can consider at any time a ``virtual''
measurement of an arbitrary magnitude $A$. The
theory will be consistent with orthodox Quantum Mechanics
if the conditional amplitude and probability distribution for any other magnitude $B$
 in the (real) state before the virtual measurement   matches
 the one of the (virtual) quantum state 
for the projection rule applied according to  the hypothetical  output. 

It is important to distinguish between virtual and real measurements. The first one is used to formally derive a distribution of probabilities without physical effects on the system, while in the second case there is real interaction, and therefore physical effects. What is forbidden at individual systems, simultaneous or 
consistent measurement of non commuting magnitudes, becomes 
observable in correlated states, where a real measurement in one system plays the role of virtual one in the companion.

A distribution of amplitudes  in the space of spin states, with arbitrary number of
directions of spin (obviously, all of them  non commuting), was given in
\cite{mio}. In the continuum limit, we consider a measurable function $f: S^2
\to \{+,-\}$, skew $f(-{\bf n}) = - f({\bf n})$,  representing the hidden state
of spin; $f({\bf n})$ is the $\pi _{\bf n} \in \{ +, -\}$ projection of the label $\lambda \equiv f$,
the result of an  hypothetical measurement in
$\bf n$ direction, and $||f>$ represents the elementary labelled state.  The amplitude of
probability for $||f>$ is

\[
\Psi (f) = \int d\Omega f({\bf n}) N
\]
where $N = ({\bf n}\cdot {\bf i}) I + ({\bf n}\cdot {\bf j}) J +  ({\bf n}\cdot
{\bf k}) K$ is a quaternion without real component.
An orthodox quantum state $|{\bf n}_0>$, with positive spin in direction ${\bf n}_0$, is
represented by 

\[
||{\bf n}_0> = \int _{f({\bf n}_0)=1} \hskip -0.3 truecm D f \,\,\Psi (f) ||f>
\]
i.e., the family of all states $||{\bf n}_0, f>$  fulfilling the condition
$f({\bf n}_0)=1$, and its associated distribution of amplitudes $\Psi (f)$.

Similarly, a solution in phase space for a point like spinless particle, 
alternative to the one given in \cite{mio},
is

\[
||S> = \int dx \int dp \,\, \Psi (x) \xi (p) e^{\frac {i}{\hbar} (p x_0 - x
p_0)} |x>'\times |p>'
\]
$(x_0,p_0)$ is a re\-fe\-ren\-ce value
at which $\Psi (x_0) \neq 0 \neq \xi (p_0)$, and we have taken advantage of
the gauge freedom, even after normalisation,
to modify the phase of an orthonormal basis of eigenstates, 
$|x>' = exp (i p_0 x /\hbar)|x>$ and $|p>' = exp (- i p x_0 /\hbar)|p>$. Trace over the second 
component 

\[
|S>  = \Psi (x_o) \int dx  \Psi (x) e^{ \frac {i}{\hbar} p_0 x} |x>' 
\]
reproduces the quantum state ray.

\section{Correlated systems}
\label{correlated}

Measurement is an interaction, a usual Hamiltonian interaction. Measurement is
an interaction in which
the initial and a family of final states for one of the interacting systems, the
measurement apparatus, can be distinguished. Then, any interaction is a
potential measurement, if we were able to distinguish among states of one of the
interacting systems. Being able to distinguish among states is a possible definition of
macroscopic state, and there is nothing fundamental on it. In fact, it evolves
with the available technology. and we can now distinguish states that were
indistinguishable in the past.

There are three ways in which we can get (necessarily  partial) information
about the state of a system, the position of the label. First, by direct
interaction with the particle, direct measurement. Second, by indirect
interaction with 
other components of the accompanying wave, indirect measurement, when some
virtual paths of the particle 
are discarded but there is no direct interaction of the particle with the
obstacle or detector. In both cases the state of the (compound particle plus
wave) system is unavoidably modified by interaction with particle or wave. In
indirect measurement, the label does not jump because it is attached to the
unperturbed particle state. But the new labelled state has less
(or shifted) wave components.

Third, we can get information on the label in a system ${\cal  S}_a$ through measurement 
on another system ${\cal  S}_b$  that has interacted in the past with ${\cal  S}_a$,
and some correlation between variables of both systems has been established. In
this case, we get  information ``without in any way disturbing the
system'' \cite{EPR} ${\cal  S}_a$, i.e., its individual quantum state does not change because
of measurement on ${\cal  S}_b$, no matter if that measurement event happens in
the past, future or with spatial separation with respect to a
 measurement on the system ${\cal  S}_a$. Of course, if there is an
instrumental arrangement in such  a way that outputs of
measurement on ${\cal  S}_b$
generate some  planned physical reaction  reaching ${\cal  S}_a$ before it is
measured, the system could change of state through a real interaction.
But projection of state  is 
applied to  ${\cal  S}_b$ because of interaction with its measurement apparatus, and it will have
in general localised effects, but not over ${\cal  S}_a$.

Let us consider two quantum systems ${\cal S}_a$ and ${\cal S}_b$, and two
correlated magnitudes $B_a$ with $B_b$ because of a
previous interaction between them. Without loss of generality, we consider that the
state of the compound system is an eigenvector of $B_a + B_b$ 
with null eigenvalue, $(B_a + B_b) |S_T> = 0$.

In a  basis of eigenvectors, 

\[
|S_T> = \sum _j z^B_j |b^a_j> \times |b^b_j = - b^a_j>
\]
in the product Hilbert space. $|S_T>$ is not of rank one, and  a change of state
(e.g., through measurement, but also through other interactions) in 
system ${\cal S}_b$ will unavoidably modify the state and table of relative frequencies for 
subsequent measurements over system ${\cal S}_a$.

With the introduction of the label, we can replace the previous entanglement by
a $\lambda$ correlation.
Systems ${\cal S}_a$ and ${\cal S}_b$ have each a quantum state
$|S_a> = \sum _j z^B_j |b^a_j>$ and $|S_b> = \sum _j z^B_j  |b^b_j>$ describing 
amplitudes and  relative frequencies, but correlated elements of
 $|S_a>$ and $|S_b>$ have also a particular label 
$(|S_a, \lambda _a>$, $|S_b, \lambda _b>)$, such
that $\pi _B( \lambda _a ) = - \pi _B(\lambda _b)$, a condition that appears at the initial interaction event.
Then,   measurements of  magnitude $B$ on a jointly generated pair
will present perfect correlation. 

In the space of spin states, the  total null spin
for particles ${\cal S}_a$ and ${\cal S}_b$ is prescribed through correlation $|S_a, \lambda
_a = f>$ and $|S_b, \lambda _b = - f>$. A statistical sample of measurements on both
particles in arbitrarily chosen directions reproduces the quantum correlation  without  projection of state. 

Let us consider another magnitude $A$ of ${\cal S}_a$. If magnitude $A$ is measured in ${\cal S}_a$, and no measurement in ${\cal S}_b$ is performed, we can ignore the correlation (i.e., trace over the second component), and
the distribution of probability is 

\[
P(a_i) = |\sum _j z^B_j <a_i|b^a_j>|^2
\]
in both the standard theory and the theory with label.

On the contrary, if  magnitude $B$ is measured in ${\cal S}_b$, and in a future  event
magnitude $A$ is measured in ${\cal S}_a$, according to the standard projection rule (which applies to 
correlated systems because of the entanglement representation) it is equivalent to performing both measurements in ${\cal S}_a$ in the exposed order. The marginal probability for $a_i$ is 

\[
P'(a_i) = \sum _j P(b_j)P(a_i | b_j) = \sum _j |z^B_j <a_i | b_j>|^2
\]

In the theory with label, measurement of $B$ in ${\cal S}_b$ is understood as a ``virtual'' measurement, i.e., just information, about the value $\pi _B(\lambda _a)$. Without projection rule, the distribution of probability for 
$a_i$ remains $P(a_i)$. We can calculate the correlation with measurement outputs in  ${\cal S}_b$,

\[
P(b_j|a_i) = |<b_j|a_i>|^2  \quad {\rm or} \quad 
P(b_j|a_i) = \frac {|Z_{ i j}|^2}{\sum _{j'}|Z_{i j'}|^2}
\]
when using joint probability distributions.

A {\it gendaken} experiment where both theories have different predictions is
a two slit experiment with particle ${\cal S}_a$, and a second system ${\cal S}_b$ 
correlated with the particle
slit variable $L$ or $R$. Both systems are generated at the origin and move in
opposite directions; position (or some other variable) of ${\cal S}_b$ is supposed
perfectly correlated with 
${\cal S}_a$ going though $L$ or $R$ slit (in an ideal case) \footnote{A
statistical sample of  measurements can previously check the correlation through joint
measurement of ${\cal S}_b$ variable and ${\cal S}_a$ position  
just behind the slits.}.  An early measurement event of ${\cal S}_b$
 can lie in the past of the
arrival event of ${\cal S}_a$ to the final screen, or even in the past of its arrival to
the  slits screen. Knowledge of the 
 $L$--$R$ variable  before ${\cal S}_a$ reaches the final screen  gives
way to projection of its state in the usual treatment, therefore, disappearance of the diffraction pattern. On the other side, if ${\cal S}_b$ variable is not
measured (or if it is measured in the future of the
arrival event of ${\cal S}_a$ to the screen), the diffraction pattern is preserved. There is
ambiguity about the predicted behaviour (diffraction pattern or not) of ${\cal S}_a$ when there
is spatial separation between both measurement events. Let us   fix the ${\cal S}_a$ measurement event; 
 initial and final boundary points, of the interval of
the ${\cal S}_b$ trajectory  in which there is 
spatial separation with ${\cal S}_a$ measurement event, do not match. There is no physical argument to select one or the other behaviour, i.e., to extrapolate one or the other boundary behaviours onto the spatial separation interval. \footnote{At particular cases in which both boundary behaviours coincide, there is no problem to 
interpolate it onto the spatial separation interval}

In the theory with label, the knowledge of the slit variable is obtained without
in any way perturbing the system,
and the diffraction pattern is always preserved, no matter the temporal relation with
the other particle measurement. That is, even for very early ${\cal S}_b$ measurements, when
we know in advance which slit is going to go ${\cal S}_a$ through, its individual state is not
perturbed. There is no
ambiguity in this prediction, the other measurement event is irrelevant for the  behaviour of ${\cal S}_a$.
There is also a contradiction with the standard prediction 
for early measurements of ${\cal S}_b$.

More specifically, with $\Psi ({\bf r}_i, L)$ and $\Psi ({\bf r}_i,R)$ denoting
the amplitudes at position 
${\bf r}_i$ of the screen for waves from $L$ and $R$ slit respectively, the 
total distribution 
on the screen is $|\Psi ({\bf r}_i, L) + \Psi ({\bf r}_i, R)|^2$ (diffraction pattern);
the conditional distribution, 
for  particles arriving to ${\bf r}_i$, between  labels  $L$ and $R$ (information obtained through 
measurements of ${\cal S}_b$) is 

\[
P(L|{\bf r}_i) = \frac {|\Psi ({\bf r}_i, L)|^2}{|\Psi ({\bf r}_i, L)|^2 + |\Psi
({\bf r}_i, R)|^2}
\quad 
P(R|{\bf r}_i) = \frac {|\Psi ({\bf r}_i, R)|^2}{|\Psi ({\bf r}_i, L)|^2 + |\Psi
({\bf r}_i, R)|^2}
\]

\section{Summary and outlook}
\label{outlook}

Two observable predictions have been obtained from the Quantum Mechanics with label: a distribution of
probabilities without projection of state for correlated systems, and existence of an additional physical system, the de Broglie wave, accompanying the particle even when isolated (so, an \ae ther). 

The first one solves an ambiguity of the standard Quantum Mechanics that appears, in generic cases, at spatially separated measurements of entangled systems, and contradicts the standard theory when the orthodox projection rule should be applied for well established time order of measurements.  Some additional interpretative issues are: non local 
interaction does not happen in this theory; measurement is a Hamiltonian, reversible interaction; quantum systems have (hidden) values of non commuting magnitudes simultaneously. However,
the alternative theory of Quantum Mechanics with label becomes, at the end of the day, a single modification of the projection rule for practical purposes: it is not applied to the state of a system when measurement is performed in a correlated system.

The de Broglie wave has perhaps higher relevance. Introduced to preserve the action reaction principle in indirect measurements, its hypothetical existence opens new insight into a bunch of issues. 
Tunnel effect, and in general energy fluctuation, would be an energy interchange between particle and accompanying wave, with strict conservation of total energy. Wave particle duality would be just wave and particle. Spontaneous emission for isolated (meta)stationary excited systems could be understood 
as induced emission of the system in a ``thermal'' bath, the surrounding \ae ther. 

The path integral formalism would have a more direct interpretation, a diffusion process of the wave.
Inspired in the path integral formalism, where paths could be grouped by final position and velocity (or momentum) of the particle, the theory with label can be applied as it is to Relativistic Quantum Mechanics
and Quantum Field Theory: particles and fields have precise, but hidden, values of non commuting physical magnitudes.  The joint use of position and velocity variables in Relativistic Quantum Mechanics could allow to
formulate it in a curved, instead of Minkowski's flat, background.

Vacuum energy effects are amply accepted as observable
 in Quantum Field Theory (Casimir energy) and Cosmology
(dark energy), at both extreme scales of length. The de Broglie wave, and perhaps also dark matter,
can be additional observable effects of the vacuum \ae ther at intermediate scales.

\section{Acknowledgements}

Financial support from research project MAT2011-22719 is acknowledged. I also
kindly ack\-now\-led\-ge helpful comments from members of the Department of
Theoretical Physics, University of Zaragoza, where I presented  some results of
this research in April 9th 2015.

\end{document}